\documentclass[%
 aip,
 amsmath,amssymb,
 reprint,%
]{revtex4-1}

\usepackage{graphicx}
\usepackage{dcolumn}
\usepackage{bm}

\usepackage[utf8]{inputenc}
\usepackage[T1]{fontenc}
\usepackage{mathptmx}
\usepackage{etoolbox}
\usepackage{physics}
\DeclareMathOperator\erfi{erfi}

\makeatletter
\def\@email#1#2{%
 \endgroup
 \patchcmd{\titleblock@produce}
  {\frontmatter@RRAPformat}
  {\frontmatter@RRAPformat{\produce@RRAP{*#1\href{mailto:#2}{#2}}}\frontmatter@RRAPformat}
  {}{}
}%
\makeatother
\begin{document}

\preprint{AIP/123-QED}

\title[Neoclassical transport and profile prediction in transport barriers]{Neoclassical transport and profile prediction in transport barriers}
\author{Silvia Trinczek}
 \email{strincze@pppl.gov}
\author{Felix I. Parra}%

\affiliation{Princeton Plasma Physics Laboratory, Princeton, NJ 08540, USA 
}
\affiliation{Department of Astrophysical Sciences, Princeton University, Princeton, NJ 08544, USA
}%

\date{\today}

\begin{abstract}
Strong gradient regions in tokamaks, such as the pedestal or internal transport barriers, are regions of reduced turbulence where neoclassical transport can play a dominant role. However, standard neoclassical transport theory assumes that the gradient length scales of density, temperature, and potential are of the order of the system size. In the pedestal, gradient length scales are much shorter and are measured to be of the order of the ion poloidal gyroradius. We present an extension of neoclassical theory that is applicable in transport barriers of large aspect ratio tokamaks. We show that particle and momentum transport are connected in such a way that a source of parallel momentum can drive a significant neoclassical ion particle flux. In strong gradient regions, density, electric potential, mean parallel flow, and ion temperature are shown to no longer be flux functions. Instead, they have a small but important poloidally varying piece that modifies the transport equations to lowest order. This introduces a nonlinearity in the transport problem through the coupling with quasineutrality that yields multiple co-existing solutions when solving for the plasma profiles. The different solutions could be connected to low and high transport states and jumps between solutions could be an indication of H-L back-transitions.

\end{abstract}

\maketitle

\section{INTRODUCTION}

Transport barriers in tokamaks are regions of reduced turbulence, where density, temperature, and electric potential change on very short length scales \cite{wagner1984, greenfieldTransportPerformanceDIIID1997}. A strong radial electric field shears the turbulence in this region and overall confinement is improved. A prominent example of a transport barrier is the pedestal, which plays a major role in high performance regimes of tokamak plasmas. The width of pedestals and of internal transport barriers (ITBs) is measured to be on the order of the ion poloidal gyroradius \cite{viezzer2018, mcdermott2009, strait1995}.\par

Neoclassical transport is usually much smaller than turbulent transport in the core of a tokamak. Transport barriers reduce turbulence sufficiently that neoclassical transport becomes the dominant transport mechanism in these regions. Experimentally, the ion heat diffusivity in pedestals in ASDEX-U\cite{viezzer2018} and in ITBs in JET\cite{tala2001} were measured to be comparable to estimates of the neoclassical ion heat diffusivity. \par

The neoclassical diffusion coefficients used in comparisons with experiments are based on standard neoclassical theory \cite{hinton1976, helander2005}, which assumes weak gradients, i.e. the gradient scale length is much larger than the ion poloidal gyroradius. This assumption is clearly violated in transport barriers. We need to extend neoclassical theory to capture ion poloidal gyroradius effects and make it applicable to regions such as transport barriers where neoclassical transport is a dominant transport mechanism. \par 

Previous work in this area was limited by assuming weak temperature gradients\cite{kagan2008,catto2011,cattoKineticEffectsTokamak2013} or weak mean parallel flow gradients\cite{shaing1992, shaing1994,shaing2012,seol2012} and importantly neglected the poloidal variation in density, temperature, and potential which forms in regions of strong gradients. Poloidal variation in pedestals has previously been observed in Alcator C-Mod \cite{theilerInboardOutboardRadial2014, churchillPoloidalAsymmetriesEdge2015} and ASDEX-U \cite{cruz-zabalaInoutChargeExchange2022}. Recent work \cite{trinczek2023, trinczek2025a} has identified a limit in which strong gradients can be included in neoclassical theory for large aspect ratio tokamaks. Ref.~\onlinecite{trinczek2023,trinczek2025a} extended neoclassical theory to regions where the density, temperature, mean parallel flow, and potential gradients are on the order of the ion poloidal gyroradius and poloidal variation was kept, calculated, and explained. It was shown that the poloidal variation is small in aspect ratio but strong enough to modify transport equations to lowest order. \par

In this article, we will show how the work of Ref.~\onlinecite{trinczek2023, trinczek2025a} is fundamentally different from standard neoclassical theory in its transport properties, and how the poloidal variation is crucial as it introduces a nonlinearity that yields multiple co-existing solutions to the transport equations with different plasma profiles. A jump between different solutions could explain H-L back-transitions.   \par
In Section \ref{sec:SGNT}, we define the ordering in which strong gradient neoclassical theory is applicable and how it relates to weak gradient neoclassical theory and finite orbit width effects. In Section \ref{sec:Momentum}, we discuss particle and momentum transport and how they are related in strong gradient regions. The existence of a parallel momentum source can change the size of the ion neoclassical particle flux by orders of magnitude. We then turn to the problem of profile prediction for given sources and boundary conditions in Section \ref{sec:Profile}. The coupling of quasineutrality to the transport relations through the poloidal variation of the electric potential introduces nonlinearities that give multiple co-existing solutions to the transport problem. A summary of our results is presented in Section \ref{sec:Conclusion}.

\section{STRONG GRADIENT NEOCLASSICAL THEORY}\label{sec:SGNT}

Standard neoclassical theory assumes sufficiently weak gradients such that finite poloidal gyroradius effects can be neglected \cite{hinton1976, helander2005}. Here, we give an overview of how Ref. ~\onlinecite{trinczek2023,trinczek2025a} extend neoclassical theory to include these poloidal gyroradius effects and how the result is different from weak gradient neoclassical theory. \par

In weak gradient neoclassical theory, the minor radius $a$, gradient length scale $L_{n,T,\Phi}$ and trapped particle orbit width $w_b$ are ordered such that
\begin{equation}\label{Core}
    a\sim L_{n,T,\Phi}\gg w_b.
\end{equation}
The gradient length scale of a quantity $Q$ is defined as $L_Q=\abs{\nabla \log Q}^{-1}$. The plasma profiles are relatively flat and many orbit widths fit within one gradient length scale. The ordering \eqref{Core} is clearly not applicable in transport barriers where $a\gg L_{n,T,\Phi}\sim \rho_p$. Here, $\rho_p$ is the ion poloidal Larmor radius. \par 

Finite orbit width effects \cite{porcelli1994, gorelenkov1999, hedin2002, imada2019} have been observed when 
\begin{equation}\label{FOW}
    a\gg L_{n,T,\Phi}\sim w_b.
\end{equation}
Here, trapped particle orbits can span the entire transport barrier width. A lot of interesting effects such as orbit loss \cite{Chankin1993}and mode stabilization effects\cite{gorelenkov1999} are based on finite sized drift orbits. However, the analytical treatment of this regime is very difficult. \par
Ref. ~\onlinecite{trinczek2023, trinczek2025a} introduced the concept of strong gradient neoclassical theory to extend neoclassical theory into transport barriers. The strong gradient regime is an intermediate regime between \eqref{Core} and \eqref{FOW} and does not include finite orbit width effects. Strong gradient neoclassical theory is only applicable in large aspect ratio tokamaks as it relies on a large aspect ratio expansion. The trapped particle orbit widths scale as $w_b\sim\sqrt{\epsilon} \rho_p$, where $\epsilon\equiv a/R$ is the inverse aspect ratio and $R$ is the major radius. When the aspect ratio is large, there is a scale separation between the orbit width and the poloidal gyroradius. Strong gradient neoclassical theory is defined by
\begin{equation}\label{strong gradient}
    a\gg L_{n,T,\Phi}\sim\rho_p\gg w_b\sim \sqrt{\epsilon}\rho_p\quad \text{and}\quad \epsilon\ll1.
\end{equation}
In this regime, there are still several orbit widths across the gradient length scale. Thus, orbits are still relatively slim and trapped particles do not span the entire transport barrier. Orbit loss is not captured by this ordering. However, the gradient scale length is still significantly smaller than the minor radius, unlike in weak gradient neoclassical theory. The advantage of \eqref{strong gradient}
is that an analytical treatment of the neoclassical transport is possible. With the approach in Ref.~\onlinecite{trinczek2023,trinczek2025a}, we can analyze the physical effects that arise when the gradient length scale is increased beyond the minor radius scale. The resulting strong gradient transport modifications would also be present for gradient scale lengths comparable to the trapped particle orbit width.\par
Another way to distinguish the strong gradient regime from the weak gradient regime is by comparing the quantity $\rho_\ast\equiv \rho/L_{n,T,\Phi}$ in the weak gradient ordering \eqref{Core} and strong gradient ordering \eqref{strong gradient}. For strong gradients $\rho_\ast\sim \rho/\rho_p\sim\epsilon$ and for weak gradients $\rho_\ast\sim\rho/a\ll\epsilon$. In the strong gradient limit, $\rho_\ast$ is of the same order as $\epsilon$ whereas in the weak gradient limit $\rho_\ast\ll \epsilon$. \par
The fundamental expansion in large aspect ratio for strong gradient neoclassical theory allows for an analytical treatment of neoclassical transport in this regime thanks to two key elements. The first aspect is the aforementioned slim orbit width. The distribution functions of ions and electrons are close to Maxwellians and the transport properties are the consequence of small corrections to these Maxwellians. The second benefit of a large aspect ratio expansion is the scale separation between the gradient length scale $L_{n,T,\Phi}\sim\rho_p$ and the Larmor radius $\rho$ because $\rho_p/\rho\sim \epsilon^{-1}\gg 1$. Thus, drift kinetics is valid in this limit and finite Larmor radius effects can be neglected. \par

Conceptually, strong gradient effects include a shift in the trapped particle region to finite parallel velocity\cite{shaing1992, kagan2008}. Particles bounce whenever their poloidal velocity goes to zero. The poloidal velocity $\dot\theta$ of a particle has two components; the poloidal projection of the parallel velocity and the poloidal projection of the $E\times B$-drift,
\begin{multline}\label{thetadot}
    \dot\theta=(v_\parallel\bm{\hat{b}}+\bm{v}_{E\times B})\cdot \nabla \theta=\left(v_\parallel+\frac{cI}{B}\pdv{\Phi}{\psi}\right)\bm{\hat{b}}\cdot \nabla \theta\\
    \equiv (v_\parallel +u)\bm{\hat{b}}\cdot \nabla\theta.
\end{multline}
Here, we introduced the velocity
\begin{equation}
    u\equiv\frac{cI}{B}\pdv{\Phi}{\psi},
\end{equation}
the parallel velocity $v_\parallel$, the magnetic field $\bm B=I\nabla\zeta+\nabla\zeta\times\nabla\psi$, the toroidal angle $\zeta$, the poloidal flux $2\pi\psi$, the magnetic field strength $B=\abs{\bm B}$, the direction of the magnetic field $\bm{\hat{b}}=\bm B/B$, the $E\times B$-drift $\bm{v}_{E\times B}$, the poloidal angle $\theta$, the speed of light $c$ and the electric potential $\Phi$.
In the weak gradient limit, the gradient of the electric potential is small and hence $u\ll v_\parallel\sim v_{th}$, where $v_{th}$ is the thermal speed of the ions. In the strong gradient limit $u\sim v_{th}$ and both terms in \eqref{thetadot} can compete. So, trapped particles are particles for which the combination of velocities $v_\parallel+u\sim\textit{O}(\sqrt{\epsilon}v_{th})$ is small. The trapped region shifts to the finite parallel velocity $v_\parallel\simeq -u$. \par
\begin{figure}
    \centering
    \includegraphics[width=\linewidth]{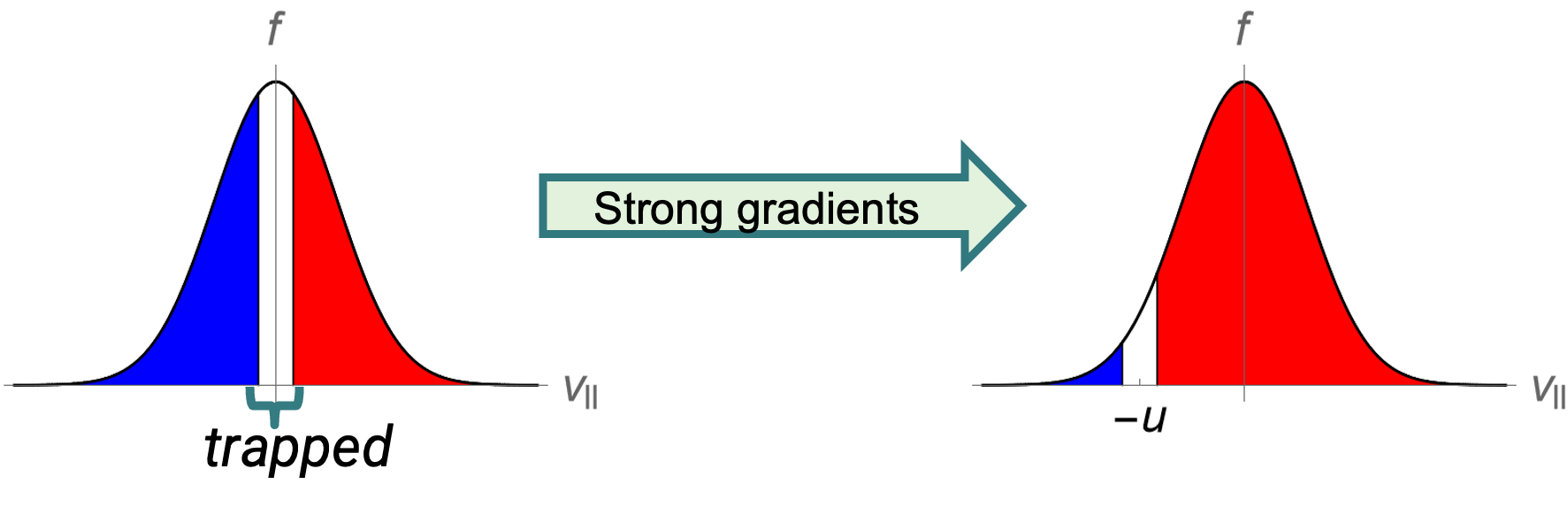}
    \caption{The trapped region in the weak gradient limit is located around $v_\parallel\simeq0$ and gets shifted to $v_\parallel\simeq-u$ in the strong gradient limit\cite{shaing1992, kagan2008}. The number of co- (red) and counter-passing (blue) particles is equal in the weak gradient limit and unequal for $V_\parallel\neq-u$ in the strong gradient limit.}
    \label{fig:Shifted}
\end{figure}

The shift in the trapped particle region introduces an asymmetry in the freely--passing particle orbits\cite{trinczek2023}. In the weak gradient limit, trapped particles are located around $v_\parallel\sim \textit{O}(\sqrt{\epsilon }v_{th})$, see Fig. \ref{fig:Shifted}. The total distribution function is close to a Maxwellian with vanishing parallel flow. Thus, the same number of co- and counter-passing particles exists. When the trapped--barely passing region is shifted to finite parallel velocity $v_{\parallel}\simeq-u$, the freely--passing particle numbers are no longer symmetric, unless $V_\parallel\simeq-u$, where $V_\parallel$ is the ion mean parallel flow. This asymmetry in freely--passing particle number is one of the four mechanisms that cause poloidal asymmetry in strong gradient regions. See Ref.~\onlinecite{trinczek2025a} for a more detailed discussion of the poloidal asymmetry in strong gradient neoclassical theory.\par
The poloidal variation in density, electric potential, flow, and ion temperature is an important result of strong gradient neoclassical theory. Ref.\onlinecite{trinczek2023} showed that the part of the electric potential $\Phi=\phi(\psi)+\phi_\theta(\psi,\theta)$ that depends on the poloidal angle is small, $\phi_\theta/\phi\sim\epsilon$. As we have pointed out above, the origin of $\phi_\theta$ was discussed in detail in Ref.\onlinecite{trinczek2025a}. The asymmetry in the banana regime is purely in-out asymmetric but a combination of in-out and up-down asymmetry forms in the plateau regime. For the remainder of this article, we will focus on the banana regime and comment on the applicability of the results in the plateau regime in Section \ref{sec:Conclusion}.\par 
For concentric circular flux surfaces, the poloidal variation in the banana regime can be written as $\phi_\theta=\phi_c(\psi)\cos\theta$, where $\theta$ is the usual poloidal angle. Importantly, although the amplitude of the poloidal variation of the electric potential $\phi_c$ is small in $\epsilon$, it is large enough to modify transport equations to lowest order. Indeed, trapped particles dominate neoclassical transport, and an electrostatic potential with a poloidal dependence can electrostatically trap and de-trap particles. A modification to the electric potential of the same size as the poloidal variation of the magnetic field, gives an order unity change in the trapping condition and thus of the trapped particle population. The neoclassical transport, which is governed by trapped particles, is modified to lowest order, too.\par

Following the approach outlined in Ref.~\onlinecite{trinczek2023}, one can find the ion neoclassical particle flux in the banana regime,
\begin{eqnarray} \label{Gamma}
    \Gamma_i^{\text{neo}}=-1.102\sqrt{\frac{r}{R}}\frac{\nu I^2  p}{\abs{S}^{3/2}m_i\Omega^2}\Bigg\lbrace \bigg[\frac{1}{p}\pdv{p}{\psi}-\frac{m_i(u+V_{\parallel})}{T}\nonumber\\
    \times \left(\pdv{V_{\parallel }}{\psi}-\frac{\Omega}{I}\right)\bigg]G_1(u,V_\parallel,\phi_c) -1.17\frac{1}{T}\pdv{T}{\psi} G_2(u,V_\parallel,\phi_c)\Bigg\rbrace,
\end{eqnarray}
where $G_1$ and $G_2$ are modifications of the diffusivities introduced in Ref.\onlinecite{trinczek2023} and listed for completeness in Appendix \ref{sec:Appendix} in \eqref{G1} and \eqref{G2}. Here,
\begin{equation}
    \nu=\frac{4\sqrt{\pi}Z^4e^4n\log\Lambda}{3T^{3/2}m^{1/2}}
\end{equation}
is the collision frequency, $n$ is the ion density, $T$ is the ion temperature, $p=nT$ is the ion pressure, $m_i$ is the ion mass, $\Omega$ is the ion Larmor frequency, $Ze$ is the ion charge, $\log\Lambda$ is the Coulomb logarithm and $S=1+(cI^2/B\Omega)(\partial^2\Phi/\partial\psi^2)$ is the squeezing factor\cite{shaing1992}. As announced above, the small poloidal variation of the potential modifies the radial fluxes: it is an argument of $G_1$ and $G_2$. \par
The ion neoclassical energy flux in the banana regime is
\begin{eqnarray}\label{Q}
     Q_i^{\text{neo}}=\frac{m_iu^2}{2}\Gamma_i^{\text{neo}}-1.463\sqrt{\frac{r}{R}}\frac{\nu I^2 p T}{\abs{S}^{3/2}m_i\Omega^2}\Bigg\lbrace\bigg[\frac{1}{p}\pdv{p}{\psi}\nonumber \\
     -\frac{m_i(u+V_{\parallel })}{T}\left(\pdv{V_{\parallel }}{\psi}-\frac{\Omega}{I}\right)\bigg]H_1(u,V_\parallel,\phi_c)\nonumber\\
        -0.25\frac{1}{T}\pdv{T}{\psi} H_2(u,V_\parallel,\phi_c)\Bigg\rbrace,
\end{eqnarray}
where the functions $H_1$ and $H_2$ are defined in Ref.~\onlinecite{trinczek2023} and stated in Appendix \ref{sec:Appendix} in \eqref{H1} and \eqref{H2}. Again, the modifications to the diffusivites $H_1$ and $H_2$ depend on the amplitude of the poloidal variation of the potential. The electron neoclassical particle and energy flux as derived in Ref.~\onlinecite{trinczek2025a} are similarly modified by functions that depend on $\phi_c$, but there is no dependence on the squeezing factor.\par
The relation of the ion neoclassical particle flux \eqref{Gamma} is the central equation in this paper, both in the context of particle and momentum transport in Section \ref{sec:Momentum}, as well as for plasma profile prediction in Section \ref{sec:Profile}.

\section{PARTICLE AND MOMENTUM TRANSPORT}\label{sec:Momentum}
The particle transport in strong gradient regions can only be understood in combination with the transport of parallel momentum. Their connection is described by the ion parallel momentum equation\cite{trinczek2023},
\begin{equation}\label{momentum}
    \pdv{}{\psi}\left(-m_iu\Gamma_i^{\text{neo}}\right)-\frac{m_i\Omega}{I}\Gamma_i^{\text{neo}}=\gamma.
\end{equation}
Here, $\gamma$ is a volumetric source of parallel momentum. The first term on the left is a flux of parallel momentum and can be interpreted as the parallel momentum of trapped particles $(-m_iu)$, which is carried by the ion neoclassical particle flux $\Gamma_i^{\text{neo}}$. The second term on the left is the parallel friction force that leads to poloidal flow damping in the weak gradient limit. \par 
The first term in \eqref{momentum} is negligible in the weak gradient limit because trapped particles do not have any flux surface averaged parallel momentum ($u\simeq 0$) and the radial derivative is weak. In the absence of a sufficiently large parallel momentum source $\gamma$, the parallel momentum equation in the weak gradient limit gives that the ion neoclassical particle flux vanishes to lowest order $\Gamma_i^{\text{neo}}\simeq0$. In the strong gradient region, $u\sim v_{th}$ and the first term in \eqref{momentum} generally does not vanish.\par
The parallel momentum source $\gamma$ can be split into an external component as well as a turbulent momentum flux
\begin{equation}
    \gamma=\gamma_{\text{ext}}-\pdv{\Pi_{\parallel}^{\text{turb}}}{\psi}.
\end{equation}
The external source could be injection of momentum by, for example, neutral beam injection (NBI). Typically, NBIs are directed into the core, where most of their parallel momentum is deposited. From there, turbulence will carry the momentum radially outwards. A radial turbulent parallel momentum flux is established. As this momentum flux reaches a transport barrier, turbulence gets sheared and turbulent transport reduces. As a result, the incoming momentum can no longer be carried by turbulence, and the decaying turbulent momentum flux acts as a source of parallel momentum in the neoclassical momentum transport equation. In this way, decaying turbulence can act as a source of parallel momentum in transport barriers. \par
Whether there is a source of parallel momentum distinguishes two distinct scenarios for the ion neoclassical particle flux.

\subsection{Without a parallel momentum source}
The ion neoclassical particle flux in the core is vanishingly small because it is limited by how fast electrons can move. In the absence of any parallel momentum source, i.e. $\gamma=0$, \eqref{momentum} gives
\begin{equation}\label{Gamma from momentum}
    \Gamma_i^{\text{neo}}\propto \exp\left(-\int\mathrm{d}\psi\:\frac{\Omega}{I}\frac{S}{u}\right).
\end{equation}
In pedestals, typically $u,S >0$, and the particle flux is small at the top of the pedestal because it is small in the core. Thus, \eqref{Gamma from momentum} gives $\Gamma_i^{\text{neo}}\rightarrow 0$. The ion neoclassical particle flux vanishes to lowest order in the absence of any parallel momentum sources. This implies that the ion particle flux must mostly be carried by turbulence in transport barriers without parallel momentum input. \par
In this scenario, \eqref{Gamma} gives a condition to determine the radial electric field $E_r$ by setting $\Gamma_i^{\text{neo}}\simeq0$,
\begin{eqnarray}\label{Gamma0}
    0=\bigg[\frac{1}{p}\pdv{p}{\psi}-\frac{m_i(u+V_{\parallel})}{T}
     \left(\pdv{V_{\parallel }}{\psi}-\frac{\Omega}{I}\right)\bigg]G_1(u,V_\parallel,\phi_c)\nonumber\\
     -1.17\frac{1}{T}\pdv{T}{\psi} G_2(u,V_\parallel,\phi_c).
\end{eqnarray}
Solving this equation for $u$ gives the radial electric field since
\begin{equation}
    E_r=-\abs{\nabla\Phi}\simeq\epsilon \frac{B}{qc}u,
\end{equation}
where $q$ is the safety factor and we used the large aspect ratio approximation in the last step.
Importantly, the solution for $E_r$ will depend on the mean parallel flow. Thus, an equation for $V_\parallel$ must be found to fully determine the radial electric field.\par
Equation \eqref{momentum} vanishes to lowest order and one has to go to higher order to calculate the parallel momentum transport across the transport barrier. This means that the momentum transport now takes place at even longer timescales because it is no longer determined by trapped particles but by passing particles. We leave the derivation of higher order parallel momentum transport in the absence of a parallel momentum source for future work. \par

In ITBs, it is possible that $S/u$ could become negative\cite{tala2001} and thus a non-vanishing ion neoclassical particle flux could be observed, but this is generally not the case in pedestals. 

\subsection{With a parallel momentum source}
If there is a parallel momentum source, such as an NBI directed at the core and decaying turbulence in the transport barrier, the neoclassical ion particle flux is not necessarily zero to lowest order. Instead, an order unity ion neoclassical particle flux is predicted. This situation is fundamentally different from what neoclassical theory predicts in the core. In the core, the neoclassical particle transport is small and most transport is carried by turbulence. The turbulence decrease in a transport barrier acts as a source of parallel momentum that drives a neoclassical ion particle flux. Now, the neoclassical particle flux is large and potentially of the order of the turbulent particle flux. If the momentum source is known, the particle flux follows from \eqref{momentum} and one can solve \eqref{Gamma} for the radial electric field. A turbulence model of momentum transport in transport barriers is required to complete this picture.

\subsection{Ambipolarity}
We want to comment on the consistency of the two scenarios, with and without parallel momentum source, with ambipolarity. Ambipolarity is the condition that the radial current in a tokamak has to vanish. In the particular regime that we are in, $L_{n,T,\Phi}\sim\rho_p$, the neoclassical transport is not necessarily intrinsically ambipolar \cite{sugamaNonlinearElectromagneticGyrokinetic1998, parraVorticityIntrinsicAmbipolarity2009, calvoLongwavelengthLimitGyrokinetics2012}. This means that $\Gamma_i^{\text{neo}}$ is not necessarily equal to the electron neoclassical particle flux $\Gamma_e^{\text{neo}}$, as is the case in weak gradient neoclassical theory. Hence, the total ion and electron particle fluxes, i.e. the combination of neoclassical and turbulent fluxes, have to balance for the radial current to vanish
\begin{equation}
\Gamma_i^{\text{neo}}+\Gamma_i^{\text{turb}}=\Gamma_e^{\text{neo}}+\Gamma_e^{\text{turb}}.
\end{equation}
We find\cite{trinczek2025a} that to lowest order in the electron-ion mass ratio, the ion neoclassical particle flux is much larger than the electron neoclassical particle flux by 
\begin{equation}
    \Gamma_i^{\text{neo}}\sim\sqrt{\frac{m_i}{m_e}}\Gamma_e^{\text{neo}}\gg\Gamma_e^{\text{neo}},
\end{equation}
where $m_e$ is the electron mass and $\Gamma_i^{\text{neo}}$ is given to lowest order in \eqref{Gamma}.
\par
If there is no parallel momentum source, the ion neoclassical particle flux vanishes to lowest order such that $\Gamma_i^{\text{neo}}\ll\Gamma_i^{\text{turb}}$ and $\Gamma_i^{\text{turb}}\sim\Gamma_e^{\text{turb}}$. Then, the turbulent contributions balance each other $\Gamma_i^{\text{turb}}=\Gamma_e^{\text{turb}}$. Since this forces the condition $\Gamma_i^{\text{neo}}\simeq0$, we call this case neoclassical ambipolarity. This is consistent with intrinsic ambipolarity in the core. \par
If there is a parallel momentum source, the ion neoclassical particle flux does not vanish to lowest order. In order to obey ambipolarity, the ion neoclassical and ion and electron turbulent particle fluxes need to balance $\Gamma_i^{\text{neo}}+\Gamma_i^{\text{turb}}\simeq\Gamma_e^{\text{turb}}$. This scenario is not neoclassically ambipolar. The electrons are carried by turbulence, whereas the ion transport is predominantly neoclassical.\par
We conclude that both in the presence and absence of a parallel momentum source, our analysis is consistent with ambipolarity as only the sum of neoclassical and turbulent fluxes need to balance.

\section{PROFILE PREDICTION}\label{sec:Profile}
The poloidal variation $\phi_\theta=\phi_c \cos\theta$ has been derived in Ref.~\onlinecite{trinczek2023} and explained physically in Ref.~\onlinecite{trinczek2025a}. It was shown that it arises due to the gradient of the mean parallel flow, the asymmetry in co- and counter-passing particle numbers, the asymmetry in the orbit widths due to the curvature drift being symmetric in $v_\parallel^2$ but not $(v_\parallel+u)^2$, and centrifugal forces. The amplitude of the poloidal variation of the electric potential is found by solving quasineutrality, assuming a Boltzmann response of the electrons. It yields an expression that conceptually looks like \cite{trinczek2023}
\begin{equation}\label{QN}
\phi_c=\phi_c\left(\pdv{n}{\psi},\pdv{T}{\psi},\pdv{V_\parallel}{\psi}\right),
\end{equation}
where the right hand side is a nonlinear function of the profile gradients. The exact form of \eqref{QN} is not relevant for the discussion below but can be found in Ref.~\onlinecite{trinczek2023} and in Appendix \ref{sec:Appendix} in \eqref{phic}.  \par 
If the density, temperature, and flow profiles are known, one can easily calculate the poloidal variation and use the amplitude $\phi_c$ together with $n$, $T$, and $V_\parallel$ in \eqref{Gamma} to calculate the neoclassical particle flux. This process is straight forward and was demonstrated successfully in Ref.\onlinecite{trinczek2023,trinczek2025a} for various profiles. \par
If one reverses this process, i.e. give the transport fluxes and determine the density, temperature, flow and radial electric field profiles, a problem arises. The dependence on $\phi_c$ of the modification functions $G_1$ and $G_2$ in \eqref{Gamma} couples \eqref{Gamma} to \eqref{QN}. The same coupling takes place in the equation for the energy transport, see \eqref{Q}. This coupling of quasineutrality and transport turns profile prediction into a nonlinear problem in the profile gradients. We will discuss two examples of how the coupling of quasineutrality and transport affects profile prediction and yields multiple co-existing solutions. \par

\subsection{Density profile prediction}\label{sec:Density}
For the first example, we imagine a situation where the particle flux, the mean parallel flow, the ion and electron temperatures, as well as the radial electric field are known and we want to solve \eqref{Gamma} for the density. Equation \eqref{Gamma} is an ODE for the density and to solve this problem numerically, we need to set a boundary value $n^{(0)}$ at the top of the pedestal and a step size $\Delta \psi$. At each position $\psi^{(j)}$, we solve \eqref{Gamma} for $(\partial n/\partial \psi)^{(j)}$ and find the density at the radial position $\psi^{(j+1)}=\psi^{(j)}+\Delta\psi$,
\begin{equation}
    n^{(j+1)}=n^{(j)}+\Delta\psi\left(\pdv{n}{\psi}\right)^{(j)}.
\end{equation}
In principle, this should give a solution for the density profile across the pedestal for a specified boundary value $n^{(0)}$. However, the solutions to \eqref{Gamma} are very sensitive to the sources and boundary conditions. \par
\begin{figure}
    \centering
    \includegraphics[width=0.85\linewidth]{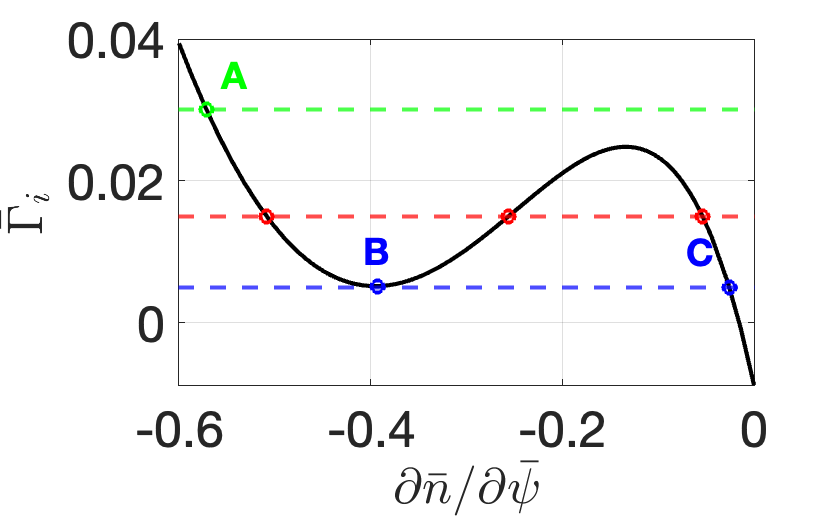}
    \caption{The circled values are the solutions to $\Gamma_i^{\text{neo}}=0.03$, $0.015$ and $0.005$ in \eqref{Gamma} for $\bar n=1$, $\bar T=0.78$, $\bar T_e=1.2$, $\partial \bar T/\partial\bar \psi=-0.72$, $\bar V=-0.75$, $\bar u=0.67$, $\partial \bar V/\partial\bar\psi=0.7$, $S=1$ and $Z=1$. There exist up to three possible solutions. The points A, B and C could correspond to a H-L back-transition.}
    \label{fig:density}
\end{figure}

An example of solving \eqref{Gamma} for the density gradient is shown in Fig. \ref{fig:density}. In this example, we solve \eqref{Gamma} for the density gradient and find up to three different solutions depending on the value of the source term $\Gamma_i^{\text{neo}}$. Small changes in $\Gamma_i^{\text{neo}}$ can lead to jumps in the solution for the gradient. Here, we normalized our quantities such that
\begin{eqnarray}
    \bar{T}\equiv\frac{T}{T_{0}},\quad \bar{n}\equiv\frac{n}{n_{0}},\quad\bar{u}\equiv\sqrt{\frac{m_i}{2T_0}}u,\quad\bar{V}\equiv\sqrt{\frac{m_i}{2T_0}}V_{\parallel },\nonumber\\
    \bar{T}_e\equiv\frac{T_e}{T_{0}},\quad \pdv{}{\bar{\psi}}\equiv\frac{I}{\Omega}\sqrt{\frac{2T_0}{m_i}}\pdv{}{\psi},\quad\bar{\Gamma}_i\equiv\frac{\Gamma_i^{\text{neo}}}{n_0I\sqrt{\frac{2T_0r}{m_iR}}\frac{\nu_0}{\Omega}},
\end{eqnarray}
where $n_0$ and $T_0$ are some typical reference values of density and temperature, $\nu_0\equiv\nu \bar T^{3/2}/\bar n$ and $T_e$ is the electron temperature. \par

Due to the existence of multiple solutions, both weak and strong gradient solutions are possible for the same source and boundary conditions. It is tempting to interpret the strong and weak gradient solutions as L-mode and H-mode and to draw an analogy between L-H transitions and jumps between the different solutions in Fig. \ref{fig:density}. However, we do not believe that the existence of multiple neoclassical transport states can explain L-H transitions. Neoclassical transport is only a small contribution to transport in L-modes, and thus turbulence must be involved in L-H transitions. In H-mode, however, turbulence is small, and thus neoclassical physics might be crucial to explain the H-L back-transition. 
 In the scenario of a back-transition, the plasma would start out in a low turbulence state with strong gradients such as solution A in Fig. \ref{fig:density}. Through a decrease in sourcing from $\bar\Gamma_i=0.03$ to $\bar\Gamma_i=0.005$, the gradient reduces to solution B. Decreasing the source further triggers a jump to solution C with a much weaker density gradient. One would also have to analyze the turbulent stability properties of each solution to determine if B and C actually correspond to low and high turbulent transport states. However, the existence of bifurcated states in a neoclassical transport formalism in H-mode is in principle consistent with the observation of hysteresis in L-H and H-L transitions, since different underlying physics might be responsible.

\subsection{Radial electric field prediction}\label{sec:Er}
Solving for $u$ is equivalent to finding the radial electric field. The case where we use \eqref{Gamma} to solve for the radial electric field, i.e. $u$, is special. Importantly, quasineutrality \eqref{QN} depends only on $u$ and not on $\partial u/\partial \psi$. Thus, \eqref{Gamma} is a linear function in $\partial u/\partial \psi$, which is hidden in the squeezing factor $S-1\propto\partial u/\partial \psi$. For a specified boundary value $u^{(0)}$, there is only one solution for the electric field profile. However, if $\Gamma_i^{\text{neo}}=0$, i.e. there is no source of parallel momentum, \eqref{Gamma} simplifies to \eqref{Gamma0}. The squeezing factor $S$ has disappeared from the equation and \eqref{Gamma0} is only an algebraic equation of $u$ that has multiple roots for the value of $u$ at each radial location. No boundary value $u^{(0)}$ can be specified. However, the algebraic equation \eqref{Gamma0} for $u$ still holds multiple roots that are sensitive to the exact values of the local gradients. This is fundamentally different to the example of the density profile prediction, where we found multiple possible roots for the gradient. \par
\begin{figure}
    \centering
    \includegraphics[width=0.85\linewidth]{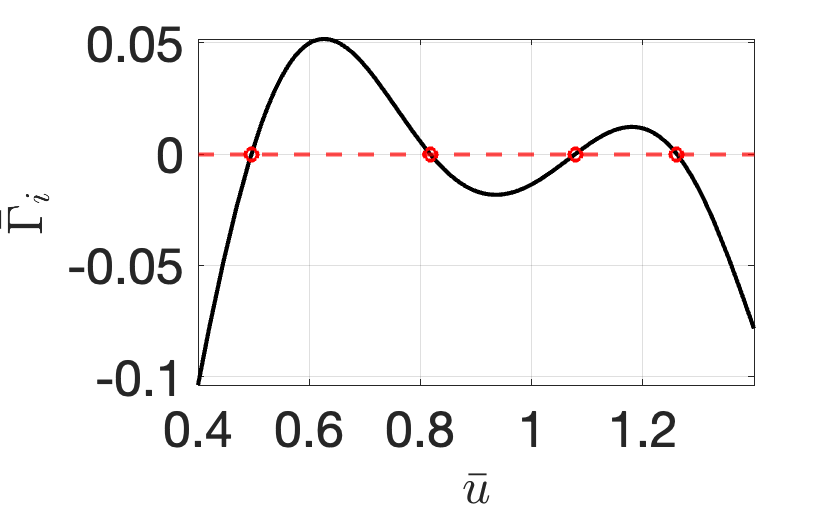}
    \caption{The circled values are the four solutions to $\Gamma_i^{\text{neo}}=0$ in \eqref{Gamma0} for $\bar n=1$, $\bar T=0.78$, $\bar T_e=1.35$, $\partial \bar T/\partial\bar \psi=-0.72$, $\partial \bar n/\partial\bar \psi=-0.4$, $\bar V=-0.75$,  $\partial \bar V/\partial\bar\psi=0.7$ and $Z=1$.}
    \label{fig:u}
\end{figure}

 
One example of solving \eqref{Gamma0} for $u$ is shown in Fig. \ref{fig:u}. In this example, there exist four consistent solutions for the local value of $u$ for given local values of density, temperature, flow, and their gradients in the absence of a source of parallel momentum. In a transport barrier scenario, the four different solutions can correspond to different depths of the radial electric field well. Although all coexisting solutions for the radial electric field are consistent with the same profiles, they result in different neoclassical energy transport, as well as different poloidal variation. 
The four possible values of $\bar u$, their corresponding neoclassical energy fluxes 
\begin{equation}
    \bar{Q}_i\equiv\frac{Q_i^{\text{neo}}}{n_0I\sqrt{\frac{2T_0r}{m_iR}}T_0\frac{\nu_0}{\Omega}},
\end{equation}
where the ion neoclassical energy flux $Q_i^{\text{neo}}$ is given in \eqref{Q}, and amplitude of poloidal variation $\bar\phi_c$ are listed in Table \ref{tab:Qphic}. We find that $\bar\phi_c$ increases with $\bar u$ but the behavior of the neoclassical energy flux is non-monotonic. The second solution yields the highest neoclassical energy flux, which is 56\% larger than the lowest neoclassical energy flux given by the fourth solution. 
\begin{table}
\caption{\label{tab:Qphic}The four different solutions for $\bar u$ in Fig. \ref{fig:u} give four different energy fluxes $\bar Q_i$ and amplitudes of the poloidal variation $\bar\phi_c$ that are all consistent with the profile values listed below Fig. \ref{fig:u}.}
\begin{ruledtabular}
\begin{tabular}{ccc}
$\bar u$&$\bar Q_i\abs{S}^{3/2}$&$\bar\phi_c$\\
\hline
0.496 & 0.629 & 1.967\\
0.818 & 0.752 & 1.986\\
1.078 & 0.624 & 2.884\\
1.261 & 0.481 & 3.828\\
\end{tabular}
\end{ruledtabular}
\end{table}

\section{CONCLUSION}\label{sec:Conclusion}
Neoclassical transport in transport barriers is modified by finite poloidal gyroradius effects. In this paper, we discussed the special ordering, derived in Ref.~\onlinecite{trinczek2023}, that makes an analytical treatment of strong gradients possible. We also showed how particle and momentum transport are connected in transport barriers and how solving for plasma profiles turns into a nonlinear problem due to the poloidal variation caused by strong gradients. \par

In a large aspect ratio tokamak, one can define a strong-gradient ordering in which the gradient scale length is comparable to the poloidal gyroradius. In the strong gradient regime $L_{n,T,\Phi}\sim\rho_p$, but the typical banana orbit width $w_b\sim\sqrt{\epsilon}\rho_p$ is smaller than $L_{n,T,\Phi}$ by $\sqrt{\epsilon}$, and there are still many orbits within a gradient length scale . This regime is defined by the ordering in \eqref{strong gradient}. Modifications to neoclassical transport in this regime are found analytically. One such modification is the shift of the trapped particle region, which introduces an asymmetry in the passing particle orbits. The poloidally varying part of the electric potential modifies the trapping condition and the neoclassical transport fluxes \eqref{Gamma} and \eqref{Q} to lowest order. \par

Parallel momentum and particle transport are related through \eqref{momentum}. The source of parallel momentum can be any form of external injection (an NBI pointed into the pedestal) or interaction with turbulence (a decaying turbulent moment flux through shear suppression in a transport barrier). We showed that the existence of a parallel momentum source can give a fundamentally different particle transport behavior. If there is no parallel momentum source, the ion neoclassical particle flux is small and we find the condition \eqref{Gamma0} for the radial electric field. A parallel momentum source can drive a strong neoclassical ion particle flux, in which case the particle transport can be predominantly neoclassical. Both cases are consistent with ambipolarity. More work is needed to understand momentum transport without parallel momentum input and the momentum source provided by a decaying turbulent momentum flux in transport barriers.

The strong gradient modification functions $G_1$ and $G_2$ depend on the poloidally varying part of the electric potential. The amplitude of the poloidal variation depends nonlinearly on the gradients of density, temperature, and flow. Solving the transport problem for these plasma profiles for a given particle flux couples quasineutrality \eqref{QN} to the particle flux equation \eqref{Gamma} through the dependence on $\phi_c$ in $G_1$ and $G_2$. The nonlinearity of the equations for the plasma profiles is unique to the strong gradient regime and does not appear in the weak gradient regime. In the example of Section \ref{sec:Density}, we showed that there exist up to three different solutions for the density gradient. The solutions are very sensitive to the prescribed flux value and do not always exist. In a second example in Section \ref{sec:Er}, we solved for the radial electric field in the case without parallel momentum source. In this case, the problem reduces to solving an algebraic equation rather than an ODE but again we find multiple possible solutions that each are consistent with the local plasma profiles but give different poloidal variation and neoclassical energy flux.

The co-existence of multiple solutions with different neoclassical transport properties and potentially different turbulent transport properties is highly suggestive of a connection to L-mode and H-mode type plasmas. A jump between different roots in, for example, Fig. \ref{fig:density} can lead to an abrupt and sharp decrease in profile gradients, similar to an H-L back-transition. Moreover, if neoclassical energy transport is assumed to be the dominant energy transport mechanism in the absence of turbulence, different consistent solutions for the radial electric field in Fig. \ref{fig:u} can lead to weaker or stronger neoclassical energy transport, see Table \ref{tab:Qphic}. However, more work is required to analyze the turbulent transport properties of the different solutions and check if they describe states of low and high turbulent transport as well as neoclassical transport.
\par

Although we only showed the transport relations in the strong gradient banana regime, the discussion also holds for the strong gradient plateau regime. The modification functions $G_1$, $G_2$, $H_1$ and $H_2$ in the neoclassical ion particle flux \eqref{Gamma} and energy flux \eqref{Q} expressions are different in the plateau regime\cite{trinczekinpreparation}. For example, there is no dependence on the squeezing factor $S$. Quasineutrality \eqref{QN} yields up-down as well as in-out asymmetry in the plateau regime, and both asymmetries modify the neoclassical fluxes. The momentum relation \eqref{momentum} remains the same and hence the discussion in Section \ref{sec:Momentum} is still applicable in the plateau regime. The dependence on the gradients is still non-linear in the expression equivalent to \eqref{QN} in the plateau regime such that there exist multiple solutions when solving for the gradients. As a result, the discussion in Section \ref{sec:Profile} mostly holds. The only difference in Section \ref{sec:Profile} is that the squeezing factor does not enter at any point such that even if there is a source of parallel momentum and $\Gamma_i^{\text{neo}}\neq0$, the relation equivalent to \eqref{Gamma} in the plateau regime is only an algebraic equation of $u$ with multiple roots and not an ODE.

\begin{acknowledgments}
This work was supported by the U.S. Department of Energy Laboratory Directed Research and Development program at the Princeton Plasma Physics Laboratory, under contract number DE-AC02-09CH11466. The United States Government retains a non-exclusive, paid-up,
irrevocable, world-wide license to publish or reproduce the published form of this
manuscript, or allow others to do so, for United States Government purposes.
\end{acknowledgments}

\section*{AUTHOR DECLARATIONS}
\subsection*{Conflict of Interest}
The authors have no conflicts to disclose.

\subsection*{Author Contributions}
\textbf{Silvia Trinczek}: Conceptualization (lead); Formal Analysis (equal); Investigation (lead); Methodology (equal); Software (lead); Visualization (lead); Writing - original draft (lead); Writing - Review \& Editing (equal). 
\textbf{Felix I. Parra}: Conceptualization (supporting); Formal Analysis (equal); Funding Acquisition (lead); Methodology (equal); Project Administration (lead); Supervision (lead); Writing - Review \& Editing (equal). 


\appendix
\section{Quasineutrality and modification functions}\label{sec:Appendix}
Here we give the expressions for the modification functions for the diffusion coefficients in the neoclassical ion particle and energy flux expressions, and the poloidal variation $\phi_c$. These results were derived in detail in Ref.~\onlinecite{trinczek2023}.\par
The modification functions $G_1$ and $G_2$ in \eqref{Gamma} are
\begin{eqnarray}\label{G1}
    G_1(u,V_\parallel,\phi_c)=7.51\int_{|(u+V_\parallel)/v_{th}|}^\infty\mathrm{d}x\:k(x,u,V_\parallel,\phi_c), 
\end{eqnarray}
and
\begin{eqnarray}\label{G2}
    G_2(u,V_\parallel,\phi_c)   \nonumber\\
    =-6.40\int_{|(u+V_\parallel)/v_{th}|}^\infty\mathrm{d}x \left(x^2-\frac{5}{2}\right)k(x,u,V_\parallel,\phi_c),
\end{eqnarray}
 where
\begin{eqnarray}\label{k}
    k(x,u,V_\parallel,\phi_c)=\sqrt{\abs{x^2+\frac{2u^2}{v_{th}^2}-\frac{Ze\phi_cR}{Tr}-\frac{(u+V_\parallel)^2}{v_{th}^2}}}\nonumber\\
    \times \bigg\lbrace\left(\frac{1}{2}-\frac{(u+V_\parallel)^2}{2v_{th}^2x^2}\right)\left[\Xi(x)-\Psi(x)\right]\nonumber\\
    +\frac{(u+V_\parallel)^2}{v_{th}^2x^2}\Psi(x)\bigg\rbrace e^{-x^2},
\end{eqnarray}
$x=|\boldsymbol{v}-V_\parallel\boldsymbol{\hat{b}}|/v_{th}$, $\Xi(x)=\text{erf}(x)=(2/\sqrt{\pi})\int_0^x\exp(-y^2)\mathrm{d}y$ is the error function and $\Psi(x)=(\Xi - x\Xi')/(2x^2)$ is the Chandrasekhar function. The functions $G_1$ and $G_2$ are normalized such that $G_1(0,0,0)=1$ and $G_2(0,0,0)=1$ in the weak gradient limit.\par

The modification functions $H_1$ and $H_2$ in \eqref{Q} are
\begin{eqnarray} \label{H1}
    H_1(u,V_\parallel,\phi_c)=5.66\int_{|(u+V_\parallel)/v_{th}|}^\infty\mathrm{d}x \left(x^2-\frac{(u+V_\parallel)^2}{v_{th}^2}\right)\nonumber\\
    \times k(x,u,V_\parallel,\phi_c)
    \end{eqnarray}
    and
    \begin{eqnarray}\label{H2}
    H_2(u,V_\parallel,\phi_c)\nonumber =-22.63 \int_{|(u+V_\parallel)/v_{th}|}^\infty\mathrm{d}x \left(x^2-\frac{(u+V_\parallel)^2}{v_{th}^2}\right)\nonumber\\
    \times\left(x^2-\frac{5}{2}\right)k(x,u,V_\parallel,\phi_c).
\end{eqnarray}

\par Quasineutrality gives the amplitude of the poloidal variation of the electric potential
\begin{widetext}
\begin{eqnarray}\label{phic}
        \Bigg\lbrace\frac{en_e}{T_e}-\frac{Z^2ne I}{T\Omega}\left[\sqrt{\frac{2T}{m_i}}J\left(\pdv{}{\psi}\ln p-\frac{3}{2}\pdv{}{\psi}\ln T\right)+\left[1-2\sqrt{\frac{m_i}{2T}}(V_\parallel+u)J\right]\right(\pdv{V_\parallel}{\psi}-\frac{\Omega}{I} -\frac{(V_\parallel+u)}{2}\pdv{}{\psi}\ln T\Bigg)\Bigg]\Bigg\rbrace\phi_c\nonumber\\
        = -2Zn\frac{r}{R}-Zn\frac{Ir}{\Omega R}\Bigg\lbrace \sqrt{\frac{2T}{m_i}}J \Bigg[\left(\frac{m_iV_\parallel^2}{T}+1\right)\left(\pdv{}{\psi}\ln p-\frac{3}{2}\pdv{}{\psi}\ln T\right)+\pdv{}{\psi}\ln T\Bigg]\nonumber\\
        +\left[1-2\sqrt{\frac{m_i}{2T}}(V_\parallel+u)J\right]\Bigg[(V_\parallel-u)\left(\pdv{}{\psi}\ln p-\frac{3}{2}\pdv{}{\psi}\ln T\right)
        +\left(\pdv{V_\parallel}{\psi}-\frac{\Omega}{I}\right)\left(\frac{m_iu^2}{T}+1-\frac{m_i(V_\parallel+u)^2 }{T}\right)\nonumber\\
        -\frac{V_\parallel+u}{2}\left(\frac{m_iV_\parallel^2}{T}+1\right)\pdv{}{\psi}\ln T\Bigg]
        +\left[1+2\frac{m_i}{2T}(V_\parallel+u)^2-4\left(\frac{m_i}{2T}\right)^{3/2}(V_\parallel+u)^3J\right]\left(\pdv{V_\parallel}{\psi}-\frac{\Omega}{I}+\frac{V_\parallel-u}{2}\pdv{}{\psi}\ln T\right)\Bigg\rbrace,       
\end{eqnarray}
\end{widetext}
where 
\begin{equation}
    J=\frac{\sqrt{\pi} }{2}\exp\left(-\frac{m_i(u+V_\parallel)^2}{2T}\right)\erfi\left(\sqrt{\frac{m_i}{2T}}(u+V_\parallel)\right)
\end{equation}
and $\erfi(x)=(2/\sqrt{\pi})\int_0^x\exp(t^2)\mathrm{d}t$.
\par 

\section*{REFERENCES}
\bibliography{MyLibrary}

\end{document}